\documentclass[11pt]{article}

\newcommand{\commentstarts}{\begin{centering}
\hspace{-1pt}\vrule\vrule
\begin{minipage}[t]{0.03\linewidth}
\hspace{0.025\linewidth}
\end{minipage}
\begin{minipage}[t]{0.95\linewidth}}
\newcommand{\commentends}{\end{minipage}
\end{centering}
\vspace{7pt}
}

\newcommand{\ashriek}{\stackrel{\cdot}{|}}
\newcommand{\intoper}{I}

\usepackage{graphicx}
\usepackage{hyperref}
\usepackage{color}
\usepackage{amsmath,amssymb,amsthm}

\newcounter{Theorems}
\newcounter{Definitions}
\newcounter{Conjectures}
\begin{document} 

\begin{titlepage}
\begin{flushright}

\end{flushright}

\begin{center}
{\Large\bf $ $ \\ $ $ \\
DGLA Dg and BV formalism
}\\
\bigskip\bigskip\bigskip
{\large Andrei Mikhailov}
\\
\bigskip\bigskip
{\it Instituto de F\'{i}sica Te\'orica, Universidade Estadual Paulista\\
R. Dr. Bento Teobaldo Ferraz 271, 
Bloco II -- Barra Funda\\
CEP:01140-070 -- S\~{a}o Paulo, Brasil\\
}

\vskip 1cm
\end{center}

\begin{abstract}
  Differrential Graded Lie Algebra Dg was previously introduced in the context of current algebras.
  We show that under some conditions, the problem of constructing equivariantly closed form
  from closed invariant form is reduces to construction of a representation of Dg.
  This includes equivariant BV formalism. In particular, an analogue of intertwiner between
  Weil and Cartan models allows to clarify the general relation between integrated and unintegrated
  operators in string worldsheet theory.
\end{abstract}

\end{titlepage}

\tableofcontents

\section{Introduction}\label{Introduction}

BV formalism is a generalization of the BRST formalism, based on the mathematical theory of
odd symplectic supermanifolds. In this formalism the path integral is interpreted as
an integral of a density of weight $1/2$ over a Lagrangian submanifold. It turns out that
this ``standard'' formulation is not sufficient to describe string worldsheet theory.
One has to also consider integration over \emph{families} of Lagrangian submanifolds.
Indeed, the idea of \cite{Schwarz:2000ct},\cite{Mikhailov:2016myt},\cite{Mikhailov:2016rkp} was to interpret integration over the worldsheet metrics
as a particular case of integration over the space of gauge fixing conditions.
Varying the worldsheet metric is a particular case of varying the Lagrangian submanifold.
Taking into account the worldsheet diffeomorphism invariance requires an equivariant
version of this integration procedure.
(In a sense, worldsheet metric is not necessarily a preferred, or ``special'', object. Varying the worldsheet
    metric is just one way to build an integration cycle, there are others. The worldsheet
    diffeomorphisms, however, \emph{are} special.)
                                           
The construction of equivariant form
involves a map of some differential graded Lie algebra (DGLA)  $D\bf g$ into the algebra of functions
on the BV phase space of the string sigma-model.
To the best of our knowledge, $D\bf g$ was first introduced, or at least clearly presented,
in  \cite{Alekseev:2010gr}.
Here we will rederive some constructions of \cite{Mikhailov:2016myt},\cite{Mikhailov:2016rkp}
using an algebraic language which emphasizes the DGLA structure,
and apply some results of  \cite{Alekseev:2010gr} to the study of worldsheet vertex operators.
In a sense, $D\bf g$ is a ``universal structure''
in equivariant BV formalism, \textit{i.e.} the ``worst-case scenario'' in terms of complexity.
The construction of $D\bf g$ is a generalization of the construction
of the ``cone'' superalgebra $C\bf g$ (which is called ``supersymmetrized Lie superalgebra''
                                               in \cite{Cordes:1994fc}).

We will now briefly outline these constructions, and the results of the present paper.

\subsection{The cone of Lie superalgebra}\label{sec:IntroCone}
     
For every Lie superalgebra $\bf a$, we can define a
DGLA $C\bf a$ (the ``cone'' of $\bf a$) as follows.
We consider vector superspace $\bf a$ as a graded vector space, such
that the grade of all elements is zero. Then, we denote $s{\bf a}$ the vector space $\bf a$ with
flipped statistics  at degree $-1$.
\footnote{In our conventions, ``grade'' corresponds to the ``ghost number'';
             statistics is \emph{not} grade mod 2.}
Consider a graded vector space:
\begin{equation}
   C{\bf a} = {\bf a}\oplus s{\bf a}
   \end{equation}
where $\bf a$ is at grade zero, and $s{\bf a}$ at grade $-1$.
(The letter $s$ means ``suspension'', the standard terminology in linear algebra.)

The commutator is defined as follows. The commutator of two elements of ${\bf a}\subset {\bf a}\oplus s{\bf a}$
is the commutator of $\bf a$, the commutator of two elements of $s\bf a$ is zero,
$s{\bf a} \subset {\bf a} \oplus s{\bf a}$ is an ideal, the action of $\bf a$ on $s\bf a$
corresponds to the adjoint representation of $\bf a$. The differential $d_{C\bf a}$ is zero on $\bf a$
and maps elements of $s\bf a$ to the elements of $\bf a$, \textit{i.e.}:
$d_{C\bf a} (sx) = x$.

This construction has an important application in differential geometry. If $\bf a$ acts
on a manifold $M$, then $C{\bf a}$ acts on differential forms on $M$. The same applies
to supermanifolds and pseudo-differential forms (PDFs) on $M$. The elements of ${\bf a}\subset {\bf a} \oplus s{\bf a}$
act as Lie derivatives. For each $x\in {\bf a}$ we denote ${\cal L}\langle x\rangle$ the corresponding
Lie derivative. The elements of $s{\bf a}\subset {\bf a} \oplus s{\bf a}$ act as ``contractions''.
For $x\in {\bf a}$, the contraction will be denoted $\iota\langle x\rangle$.
(We use angular brackets $f\langle x\rangle$ when $f$ is a linear function, to highlight linear dependence on $x$.)
We have:
\begin{align} d \iota\langle x\rangle \;=\;
 &{\cal L}\langle x\rangle\label{DefCone} \\ d {\cal L}\langle x\rangle \;=\;
 &0\nonumber{} \\ [{\cal L}\langle x\rangle, {\cal L}\langle y\rangle] \;=\;
 &{\cal L}\langle [x,y]\rangle\nonumber{} \end{align}

\subsection{$D\bf g$}\label{sec:IntroDg}

The definition of $D\bf g$ is similar to the definition of $C\bf g$. Essentially, we replace
the commutative ideal  $s{\bf g}\subset C{\bf g}$ with a free Lie superalgebra of the linear space
$s^{-1}\mbox{Symm}(s^2{\bf g})$ where $\mbox{Symm}(s^2{\bf g})$ is the space of symmetric
tensors of $s^2\bf g$. 
Instead of defining  the commutators to be zero, we only require that some linear combinations of
commutators are $d_{D\bf g}$-exact. Eq. (\ref{DefCone}) is replaced with:
\begin{align} di(x) + {1\over 2}[i(x),i(x)] \;=\;
 &l\langle x\rangle\label{IntroDefDg} \\ dl\langle x\rangle \;=\;
 &0\nonumber{} \\ [l\langle x\rangle , l\langle y\rangle] \;=\;
 &l\langle [x,y] \rangle\nonumber{} \end{align}
In particular, if $i(x)$ is a linear function of $x$, then $D\bf g$ becomes $C\bf g$. In this case,
$i(x)$ becomes $\iota\langle x \rangle$ and $l\langle x\rangle$ becomes ${\cal L}\langle x \rangle$.
In general, $i(x)$ is a nonlinear function of $x$ (but $l\langle x\rangle$ remains linear).
In Section \ref{DefDg} we explain the details of the construction, and why it is
natural. We slightly generalize it, by allowing $\bf g$ to be  a  Lie superalgebra
(while in \cite{Alekseev:2010gr} it was a Lie algebra).

\subsection{String measure}\label{sec:IntroStringMeasure}

In BV formalism, to every half-density $\rho_{1/2}$ satisfying the Quantum Master Equation corresponds
a closed PDF on the space of Lagrangian submanifolds, which we denote $\Omega$
\cite{Mikhailov:2016myt},\cite{Mikhailov:2016rkp}. Besides being closed, it satisfies the following
very special property:
\begin{equation}
   d\iota\langle  x\rangle \;\Omega\;=\; \iota\langle{\Delta x}\rangle \;\Omega
   \label{SpecialPropertyOfOmega}\end{equation}
where $x\in\bf a$, $\bf a$ is the algebra of infinitesimal odd canonical transformations, 
and $\Delta$ is some differential on $\bf a$, which is associated to the half-density $\rho_{1/2}$.
This form $\Omega$ is \emph{inhomogeneous},
\textit{i.e.} does not have a definite rank. It is, generally speaking, a pseudo-differential form (PDF).
Otherwise, Eq. (\ref{SpecialPropertyOfOmega}) would not make sense.
We rederive $\Omega$ and explain its meaning as a Lie superalgebra cocycle in
Section \ref{sec:CorrelationFunctionsAreCocycle}.

\subsection{Equivariant string measure}\label{sec:IntroEquivariant}

Let ${\bf g}\subset {\bf a}$ be the algebra of vector fields on the worldsheet.
In the BV approach to string worldsheet theory,  worldsheet diffeomorphisms are symmetries of $\rho_{1/2}$,
and therefore   $\Omega$ is $\bf g$-invariant. We are interested in constructing the ${\bf g}$-equivariant version of $\Omega$.
Generally speaking, there is no good algorithm for constructing an equivariant PDF out of an invariant PDF.
But in our case, since $\Omega$ satisfies Eq. (\ref{SpecialPropertyOfOmega}), we can reduce the construction
of equvariant form to the construction of an embedding  ${D\bf g}\rightarrow {\bf a}$
--- see Section \ref{AnsatzForEquivariantForm}.

\subsection{Vertex operators}\label{sec:IntroVertexOperators}

Consider \emph{deformations} of $\Omega$. In string theory context they are called ``vertex operators''.
It is useful to consider deformations which break some of the symmetries. Typically, we insert
some operators at some points on the wordsheet, breaking the diffeomorphisms down to the subgroup preserving
that set of points. This is the ``unintegrated vertex operator''. Then, there exists an averaging
procedure which restores the symmetry group back to all diffeomorphisms.
The result of this averaging is effectively an insertion of ``integrated vertex operator'' which preserves
\emph{all} the diffeomorphisms. This relation between unintegrated and integrated vertex operators is important
in string theory.

As we show in Section \ref{IntegratedAndUnintegrated},
this averaging procedure requires an action of $D\bf g$.  Just to define the action of
symmetries, we only need $l\langle x\rangle$. But the averaging procedure,
which is needed to compute string amplitudes, does involve $i(x)$. 
In previously studied cases, such as bosonic or NSR string, $D\bf g$ reduces to $C\bf g$,
and $i(x)$ is a very simple expression. It is basically the contraction of $x$ with the ghost antifield,
\begin{equation}
   i(\xi) = \int c^{\star}_{\alpha}\xi^{\alpha} 
   \end{equation}
using the
\href{https://andreimikhailov.com/math/bv/bosonic-string/IntegrationOverFamily.html#%28part.%5F.Choice.Of.F%29}{\textbf{\textcolor{blue}{notations of \cite{Mikhailov:2016rkp}}}}
(see also Section \ref{sec:GhostNumber}).
The averaging procedure consists in this case of simply removing the ghost fields from the vertex,
and then integrating over the insertion point.
In Section \ref{IntegratedAndUnintegrated}
we derive the general formula, which is rather nontrivial and uses some intertwining
operator constructed in \cite{Alekseev:2010gr}.

At this time, we do not have concrete examples of string worldsheet theories where $D\bf g$ would
not enter only through the projection to $C{\bf g}$. It is likely that pure spinor superstring in AdS
background is an example, but we only have a partial construction \cite{Mikhailov:2017mdo}.

\subsection{Previous work}\label{sec:PreviousWork}

Previous work on equivariant BV formalism includes
\cite{Nersessian:1993me}, \cite{Nersessian:1993eq}, \cite{Nersessian:1995yt},
\cite{Getzler:2015jrr}, \cite{Getzler:2016fek}, \cite{Cattaneo:2016zrn}, \cite{Getzler:2018sbh},
\cite{Bonechi:2019dqk}.
Similar algebraic structures appeared recently (in a different context?) in \cite{Bonezzi:2019bek} and
references therein.

\section{Notations}\label{Notations}

When a function $f(x)$ depends on $x$ linearly, we will write:
\begin{equation}
   f\langle x\rangle \quad\mbox{\tt instead of}\quad f(x)
   \end{equation}
to stress linearity.
The cone of the Lie superalgebra $\bf g$ is:
\begin{equation}
   C{\bf g} = {\bf g} \stackrel{\rightarrow}{\oplus} s{\bf g}
   \end{equation}
where $\stackrel{\rightarrow}{\oplus}$ stands for semidirect sum of Lie superalgebras, with arrow pointing towards the ideal,
and $s$ is suspension, of degree $-1$. (We consider $\bf g$ as a graded vector space, all having
                                               degree zero, then $s\bf g$ all has degree $-1$.)
\footnote{
         It may seem strange to assign degree $-1$ to $s$, instead of degree $+1$.
         In string theory context, we want the grade to be the ghost number.
         At the same time, we do not want to replace $s$ with $s^{-1}$, because we want
         to agree with the notations of \cite{LodayVallette}.
         }
Throughout the paper we will follow
\href{http://irma.math.unistra.fr/~loday/PAPERS/LodayVallette.pdf}{\textbf{\textcolor{blue}{the notations}}}
of \cite{LodayVallette}.
From a vector space $V$ over a field $\bf K$ (for us ${\bf K} ={\bf R}$) we construct
an algebra $A$, which consists of tensors of $V$ modulo some quadratic relations $R\subset V\otimes V$.
The coalgebra $A^{\ashriek}$ consists of tensors of $sV$, such that the tensor product of
any pair of neighbor $V$ fall into $s^2R$. 
The cobar construction of $A^{\ashriek}$ is denoted $\Omega\left(A^{\ashriek}\right)$.

One motivation for using the formalism of quadratic algebras is technical, as it automates
keeping track of signs. 
We will translate into ``more elementary'' language of \cite{Alekseev:2010gr} in 
Section \ref{sec:FunctionsOfCurvature}.

PDF = pseudo-differential form.

Hyperlinks to web content \href{https://andreimikhailov.com/math/bv/omega/Equivariant_half-densities.html}{\textbf{\textcolor{blue}{are highlighted blue}}}.

\section{$D\bf g$}\label{DefDg}

\subsection{Definition of $D\bf g$}\label{sec:DefDg}

As far as we know, $D\bf g$ was introduced in \cite{Alekseev:2010gr}, in the context of current algebras.
We will now present it in the language of quadratic algebras.

\subsubsection{Supercommutative algebra and its dual coalgebra}\label{sec:DualAlgebraOfFunctions}

Let $V$ be a super linear space, and $T^a(V)$ be the tensor algebra generated by $V$:
\begin{equation}
   T^a(V) = \bigoplus_n V^{\otimes n}
   \end{equation}
Here the upper index $a$ implies that we consider tensors as forming an \emph{algebra}.
Let $A$ be the free super-commutative algebra generated by $V$. We define it as the subspace
of $T^a(V)$ consisting of symmetric tensors.
(To define the structure of algebra, we notice that this subspace is isomorphic to the
    factorspace by \emph{quadratiic relations} of $A$, which are antisymmetric tensors in $V\otimes V$.)

Then, the dual coalgebra $A^{\ashriek}$, by definition, the subspace of \emph{symmetric tensors} in the tensor
coalgebra:
\begin{equation}
   A^{\ashriek} \subset T^c(sV) = \bigoplus_n (sV)^{\otimes n}
   \end{equation}
(Here the upper index $c$ means that we consider it as a coalgebra.)

For example, let $a\in V,\,b\in V$ be even and $\psi\in V,\,\eta\in V$ be odd.
The following tensors belong to $A$:
\begin{equation}
   a\otimes b + b\otimes a\,,\;
   a\otimes \xi + \xi\otimes a\,,\;
   \xi\otimes\eta - \eta\otimes\xi
   \end{equation}
The bar construction is:
\begin{align}  
 &BA = T^c(s\overline{A})\nonumber{} \\  
 &d_B(sx\otimes_B sy) = (-1)^{\bar{x}} sxy\nonumber{} \end{align}
The subspace $A^{\ashriek}\subset BA$ is annihilated by $d_B$ \emph{because of relations} of $A$.
In particular, the following are elements of $A^{\ashriek}$:
\begin{equation}
   sa\otimes sb - sb\otimes sa\,,\;
   sa\otimes s\psi + s\psi \otimes sa\,,\;
   s\psi\otimes s\eta + s\eta \otimes s\psi
   \end{equation}
These are \emph{symmetric tensors} in $sV$.

To summarize, if $A$ is the algebra of symmetric tensors in $V$,
then $A^{\ashriek}$ is the coalgebra of symmetric tensors in $sV$.

\subsubsection{Case of $V =s{\bf g}$}\label{sec:DefA}

Let $\bf g$ be a Lie superalgebra. We consider it as a graded Lie superalgebra, with all elements
having grade zero. Let us apply the construction of
Section \ref{sec:DualAlgebraOfFunctions}
to $V =s{\bf g}$.

Let $A$ be the free commutative superalgebra generated by $V$, and $A^{\ashriek}$ its Koszul dual
coalgebra:
\begin{align} A\;\subset\;
 &T^a(s{\bf g})\nonumber{} \\ A^{\ashriek} \;\subset\;
 &T^c(s^2{\bf g})\nonumber{} \end{align}
At this point, we consider $s{\bf g}$ only as a supercommutative algebra.
The Lie algebra structure on $\bf g$ is forgotten. 

Let us consider the cobar construction of $A^{\ashriek}$:
\begin{equation}
   \Omega\left( A^{\ashriek}\right) = T^a\left(s^{-1}  \overline{A^{\ashriek}}\right)
   \end{equation}
Now the index $a$ in $T^a$ means that we consider tensors as forming an algebra, the free algebra.
The overline over $A^{\ashriek}$ means that we remove
the counit, see \cite{LodayVallette} for precise definitions.
We actually only need a subspace:
\begin{equation}
   \mbox{FreeLie}\left(s^{-1}  \overline{A^{\ashriek}}\right) \subset   \Omega\left( A^{\ashriek}\right)
   \end{equation}
which is generated by commutators. This is a free Lie superalgebra.
Consider the natural twisting morphism
\begin{align} \alpha \;:\;
 &A^{\ashriek} \rightarrow \Omega\left( A^{\ashriek}\right)\label{NaturalTwistingMorphism} \end{align}
(which is denoted $\iota$ in Chapter 2 of \cite{LodayVallette},
       but we reserve $\iota$ for contraction of a vector field into a form).
Its image belongs to $s^{-1}  A^{\ashriek}\subset \mbox{FreeLie}\left(s^{-1}  \overline{A^{\ashriek}}\right)$. It satisfies
the Maurer-Cartan equation;
using the notations of Chapter 2 of \cite{LodayVallette}:
\begin{equation}
   d_{\Omega} \alpha + \alpha * \alpha = 0
   \label{MCAlpha}\end{equation}
where $d_{\Omega}$ is the differential on $\Omega\left( A^{\ashriek}\right)$ induced by the coalgebra structure on $A^{\ashriek}$.
Since $A$ is a supercommutative algebra, $d_{\Omega}\left(s^{-1}  \overline{A^{\ashriek}}\right)$ actually
belongs to the subspace:
\begin{equation}
   \left(s^{-1}  \overline{A^{\ashriek}} \right)\wedge \left(s^{-1}  \overline{A^{\ashriek}} \right)
   \subset
   \left(s^{-1}  \overline{A^{\ashriek}} \right)\otimes \left(s^{-1}  \overline{A^{\ashriek}} \right)
   \end{equation}
This implies that $d_{\Omega}$ preserves the subspace:
\begin{equation}
   \mbox{FreeLie}\left(s^{-1}  \overline{A^{\ashriek}}\right)
   \subset
   \Omega\left( A^{\ashriek}\right)
   \end{equation}
(Indeed, the $H^0$ of the cobar complex is $A$, and $A$ is supercommutative;
         $d_{\Omega}$ ``kills the commutator''.)
This means that we may write ${1\over 2}[\alpha\stackrel{*}{,}\alpha]$ instead of $\alpha *\alpha$.

To summarize, $\mbox{FreeLie}\left(s^{-1}  \overline{A^{\ashriek}}\right)$ with $d_{\Omega}$ is a differential graded Lie superalgebra.

\subsubsection{Definition of $D\bf g$}\label{subsec:Dg}

Let us consider a larger space:
\begin{equation}
   D{\bf g} \;=\; {\bf g} \stackrel{\rightarrow}{\oplus} \mbox{FreeLie}\left(s^{-1}  \overline{A^{\ashriek}}\right)
   \label{SemidirectSum}\end{equation}
Here $\stackrel{\rightarrow}{\oplus}$ stands for semidirect sum of Lie superalgebras, with arrow pointing towards the ideal.
The embedding of $\bf g$ into $D\bf g$ as the first summand will be denoted $l$:
\begin{equation}
   l\;:\;{\bf g} \rightarrow D{\bf g}
   \label{EmbeddingL}\end{equation}

\paragraph{Lie superalgebra structure on $D\bf g$}\label{paragraph:CommDg}

We define the commutator of two elements of $D\bf g$ as follows:

\begin{itemize}\item 
        The commutator of two elements of $\mbox{FreeLie}\left(s^{-1}  \overline{A^{\ashriek}}\right)$ is the commutator
        of the free Lie algebra.
        \item 
        The commutator of two elements of $\bf g$ is the commutator of $\bf g$.
        \item         
        The commutator of elements of $\bf g$ and elements of $\mbox{FreeLie}\left(s^{-1}  \overline{A^{\ashriek}}\right)$
        corresponds to the natural action of $\bf g$ on $A^{\ashriek}$.
        \end{itemize}

\paragraph{Differential $d'$ of $D\bf g$}\label{paragraph:DifferrentialOfDg}

There is a natural projection:
\begin{equation}
   \pi \;:\; A^{\ashriek} \rightarrow s^2{\bf g}
   \end{equation}
annihilating all tensors with rank $\geq 1$
(\textit{i.e.} $\mbox{ker}\pi = \left(A^{\ashriek}\right)^{\geq 1}$).
We define a differential $d'$ on ${\bf g} \stackrel{\rightarrow}{\oplus} \mbox{FreeLie}\left(s^{-1} \overline{A^{\ashriek}}\right)$, in the following way.
\begin{itemize}\item We postulate that the action of $d'$ on $\bf g$ (the first term in Eq. (\ref{SemidirectSum})) be zero:
\begin{equation}
                        d'|_{\bf g} = 0
                        \label{DPrimeOnG}\end{equation}
\item It is enough to define the action of $d'$ on
              $s^{-1}  A^{\ashriek} \subset \mbox{FreeLie}\left(s^{-1}  \overline{A^{\ashriek}}\right)$. 
We put:
\begin{equation}
                        d'|_{s^{-1}  A^{\ashriek}} = s^{-1}\circ\pi \;:\; s^{-1}  A^{\ashriek} \rightarrow {\bf g}
                        \label{DPrimeOnA}\end{equation}
\end{itemize}
We then extend Eqs. (\ref{DPrimeOnG}) and (\ref{DPrimeOnA}) to the differential $d'$ of 
${\bf g}\stackrel{\rightarrow}{\oplus}\mbox{FreeLie}\left(s^{-1}  \overline{A^{\ashriek}}\right)$.

We consider  ${\bf g} \stackrel{\rightarrow}{\oplus} \mbox{FreeLie}\left(s^{-1} \overline{A^{\ashriek}}\right)$ with the differential $d_{\Omega} + d'$ which will be called $d_{D\bf g}$:
\begin{equation}
   d_{D\bf g}=d_{\Omega} + d'
   \end{equation}
The nilpotence of $d_{D\bf g}$ follows from the fact that $d_{\Omega}$ anticommutes with $d'$:
\begin{equation}
   d'd_{\Omega} + d_{\Omega} d' = 0
   \label{DsAnticommute}\end{equation}
This is proven in Section \ref{sec:AppendixNilpotence}.

\subsection{Representation as vector fields}\label{sec:RepVect}

Consider the \emph{cone} of our free Lie algebra:
\begin{align} {\cal C} \;=\;
 &C(\mbox{FreeLie}\left(s^{-1}  \overline{A^{\ashriek}}\right))\nonumber{} \\ {\cal L}\;:\;
 &\mbox{FreeLie}\left(s^{-1}  \overline{A^{\ashriek}}\right)
             \longrightarrow C(\mbox{FreeLie}\left(s^{-1}  \overline{A^{\ashriek}}\right))\nonumber{} \\ \iota\;:\;
 &\mbox{FreeLie}\left(s^{-1}  \overline{A^{\ashriek}}\right)
          \longrightarrow C(\mbox{FreeLie}\left(s^{-1}  \overline{A^{\ashriek}}\right))\nonumber{} \end{align}
and its universal enveloping algebra $U{\cal C}$. The Maurer-Cartan Eq. (\ref{MCAlpha}) implies:
\begin{align}  
 &
        (d_{\cal C} + d_{\Omega}) \exp(-\iota\circ\alpha) =
        \exp(-\iota\circ\alpha) (d_{\cal C} + d_{\Omega} + {\cal L}\circ\alpha)
        \label{DWithExp} \end{align}
This is an equation in the completion of $\mbox{Hom}\left( A^{\ashriek}\,,\,U{\cal C}\right)$.
Let $M$ be some supermanifold, and $\mbox{Vect}(M)$ the algebra of vector fields on it.
Suppose that we are given a map of linear spaces:
\begin{equation}
   s^{-1}  A^{\ashriek} \rightarrow \mbox{Vect}(M)
   \end{equation}
Such a map defines a representation of $\cal C$ in the space of pseudo-differential forms (PDFs) on $M$.
We want to project Eq. (\ref{DWithExp}) on the space of PDFs on $M$. It is not possible to do directly,
because we do not require that $d_{\Omega}$ act on PDFs.
Instead, we will use the following version of Eq. (\ref{DWithExp}):
\begin{align}  
 &
        d_M \exp(-\iota\circ\alpha) \omega =
        \exp(-\iota\circ\alpha) (d_M + \iota\circ (d_{\Omega}\alpha) + {\cal L}\circ\alpha)\omega
        \label{DMwithExpIota} \\  
 &\omega \in \mbox{Hom}( A^{\ashriek},\mbox{Fun}(\Pi T M))\nonumber{} \end{align}
for all $\omega \in \mbox{Hom}( A^{\ashriek},\mbox{Fun}(\Pi T M))$.

\subsection{Simpler notations}\label{sec:FunctionsOfCurvature}

For us $V = s{\bf g}$, and $A^{\ashriek}$ is the coalgebra
of symmetric tensors in $s^2\bf g$. Therefore, the space $\mbox{Hom}\left(A^{\ashriek},L\right)$
can be thought of as the space of formal Taylor series of functions on the superspace $s^2\bf g$
with values in a linear superspace $L$.
The subspace $\mbox{Hom}\left(\left(A^{\ashriek}\right)^n,L\right)$,
where $\left(A^{\ashriek}\right)^n$ consists of rank $n$ tensors, is the space
of $n$-th coefficients of the Taylor series. In particular, we interpret
$\mbox{Hom}( A^{\ashriek},\mbox{Fun}(\Pi T M))$ as the space of functions on the supermanifold
$s^2{\bf g} \times \Pi TM$:
\begin{equation}
   \mbox{Hom}\left(
                   A^{\ashriek},\mbox{Fun}(\Pi T M)
                   \right) \simeq \mbox{Fun} (s^2{\bf g}\times \Pi TM)
   \end{equation}
where $\simeq$ means that we are not being rigorous. We ignore the question of which functions are allowed,
\textit{i.e.} do not explain the precise meaning of $\mbox{Fun}(\ldots)$.

Let $\{e_a\}$ denote some basis in $s^2\bf g$, and
$\{F^a\}$ the dual basis in the space $s^{-2}{\bf g}^*$ of linear functions on $\bf g$:
\begin{equation}
   F^a \in \mbox{Hom}(s^2{\bf g}, {\bf K}) \quad\mbox{\tt\small for }\quad a\in \{1,\ldots,{\rm dim}{\bf g}\}
   \end{equation}
In this language:
\begin{align}  
 &\alpha \;=\; F^{a_1}\cdots F^{a_n} \; s^{-1}(e_{a_1}\otimes\cdots\otimes e_{a_n})\nonumber{} \\  
 &T^{(a_1\ldots a_n)}d_{\Omega} s^{-1}(e_{a_1}\otimes\cdots\otimes e_{a_n})\;=\;
          (-)^{a_1}T^{(a_1\ldots a_n)} s^{-1}e_{a_1}\otimes s^{-1}(e_{a_2}\otimes\cdots\otimes e_{a_n})\;+\nonumber{} \\  
 &+\;(-)^{a_1+a_2}T^{(a_1\ldots a_n)} s^{-1}(e_{a_1}\otimes e_{a_2})\otimes s^{-1}(e_{a_3}\otimes\cdots\otimes e_{a_n})+\ldots\nonumber{} \end{align}
for $T^{(a_1\ldots a_n)}$ any tensor symmetric in $a_1\ldots a_n$.
To agree with \cite{Alekseev:2010gr}, we will denote:
\begin{align}  
 &i_{a_1\ldots a_n} = s^{-1}(e_{a_1}\otimes\cdots\otimes e_{a_n})\nonumber{} \\  
 &i(F) = \alpha = F^{a_1}\cdots F^{a_n} i_{a_1\ldots a_n}\label{iF} \\  
 &l\langle F\rangle = F^a e_a \mbox{ \tt\small where } e_a\in{\bf g}\subset {\bf g} \stackrel{\rightarrow}{\oplus} \mbox{FreeLie}\left(s^{-1}  \overline{A^{\ashriek}}\right)\label{lF} \end{align}
Here the notation $l\langle F\rangle$ agrees with Eq. (\ref{EmbeddingL}).

\subsection{Ghost number}\label{sec:GhostNumber}

In our notations, if $X$ has ghost number $n$ then $sX$ has ghost number $n-1$.
In other words, $s$ \emph{lowers} ghost number. In particular, the cone of the Lie superalgebra
$\bf g$ is:
\begin{equation}
   C{\bf g} = {\bf g} \stackrel{\rightarrow}{\oplus} s{\bf g}
   \end{equation}
The generator $i_{a_1\ldots a_n}$ has ghost number $-2n+1$.
For example, in
\href{https://andreimikhailov.com/math/bv/bosonic-string/SolutionOfMasterEqn.html}{\textbf{\textcolor{blue}{bosonic string theory}}},
$i_a$ corresponds to $\{c^{\star},\_\} = {\partial\over\partial c}$
where $c^{\star}$
is the BV antifield for ghost and $\{\_,\_\}$ the odd Poisson bracket. As a mnemonic rule,
$i_{a_1\ldots a_n}$ has the same ghost number as $\{(c^{\star})^n,\_\}$. Elements of
${\bf g}$ (the first summand in ${\bf g} \stackrel{\rightarrow}{\oplus} \mbox{FreeLie}\left(s^{-1}  \overline{A^{\ashriek}}\right)$) all have ghost number zero.

\subsection{Are $F^a$ Faddev-Popov ghosts?}\label{AsFaddeevPopov}

As an ideal in $C\bf g$, the $s\bf g$ is Lie superalgebra with zero commutator. We can think of
$F^a$ as Faddeev-Popov ghosts of the BRST complex of $s\bf g$. Since the commutator is zero,
the BRST differential annihilates $F^a$.

But when we consider $D\bf g$ and not $C\bf g$, the commutator $[i_a,i_b]$ is nonzero;
we consider the free Lie superalgebra generated by $i_a$. We might have introduced new
Faddeev-Popov ghost for nested commutators, and the corresponding BRST differential.
But this is not how the construction goes. 
Instead, we introduce new generators, such as $i_{ab}$, and the differential $d_{\Omega}$
such that $di_{ab} = [i_a,i_b]$ \textit{etc}. This differential $d_{\Omega}$ acts on elements
of the Lie superalgebra, not on Faddeev-Popov ghosts.
This is \emph{not} the Faddeev-Popov construction. Notice that $i_{a_1\ldots a_n}$ gets contracted
with products of $F^a$. In this sense, we may say that we replace the Faddeev-Popov
$c^A t_A$ with nonlinear functions of $c$. We replace:
\begin{equation}
   c^A t_A + {1\over 2} f_{AB}^C c^A c^B{\partial \over \partial c^C}
   \end{equation}
with:
\begin{equation}
   \sum F^{a_1}\cdots F^{a_n} i_{a_1\cdots a_n} + d_{\Omega}
   \end{equation}

\section{$D'{\bf g}$}\label{ExtensionOfDg}

We need to also consider an extension of $D{\bf g}$, which we will call $D'{\bf g}$, which is obtained
by replacing $\mbox{FreeLie}\left(s^{-1}  \overline{A^{\ashriek}}\right)$
with $\mbox{FreeLie}(s^{-1}  A^{\ashriek})$:
\begin{equation}
   D'{\bf g} = {\bf g}\oplus \mbox{FreeLie}(s^{-1}A^{\ashriek})
   \end{equation}
In the language of Section \ref{sec:FunctionsOfCurvature}, we allow $i(0)\neq 0$, but require:
\begin{align}  
 &d_{D'\bf g} i(0) + {1\over 2} [i(0),i(0)] = 0\nonumber{} \\  
 &[x, i(0)]=0 \mbox{ \tt\small for all } x\in {\bf g}\nonumber{} \end{align}

\section{Ansatz for equivariant form}\label{AnsatzForEquivariantForm}

Suppose that the Lie superalgebra $\bf g$ is realized as vector fields on some supermanifold $M$.
This means that pseudo-differential forms on $M$ are a $\bf g$-differential module
(= representation of $(C{\bf g}, d_{C\bf g})$).
Consider the equation:
\begin{align}  
 &(d + \iota\langle F\rangle) \omega^{\tt C} = 0\label{CartanEq} \\  
 &\omega^{\tt C}\in \mbox{Fun}(s^2{\bf g}\times \Pi TM)\nonumber{} \end{align}
We can write such an equations in any $\bf g$-differential module $W$, not necessarily
PDFs on $M$.

Now suppose that $W$ is a $D'\bf g$-differential module (= representation of $CD'\bf g$).
Since $\bf g$ is embedded into $D'{\bf g}$,
we can still write the Cartan Eq. (\ref{CartanEq}):
\begin{equation}
   (d + \iota\langle l\langle F\rangle \rangle)\omega^{\tt C} = 0
   \end{equation}
where we use the notations of Eqs. (\ref{EmbeddingL}), (\ref{lF}).

Then, given any $d$-closed
form $\omega$, consider the following anstaz for a solution of Eq. (\ref{CartanEq}):
\begin{equation}
   \omega^{\tt C} = \exp(\iota\langle i(F)\rangle)\omega
   \end{equation}
where $i(F)$ is from Eq. (\ref{iF}).
We will show that under certain conditions this substitution solves Eq. (\ref{CartanEq}).

\section{$CD\bf g$}\label{SpecialCocycles}

\subsection{$D'\bf g$-differential modules}\label{DGDifferentialModules}

Suppose that $W$ is a $D'\bf g$-differential module. This means that $W$
is a representation of the Lie superalgebra $CD'\bf g$,  with the
differential $d_W$ which agrees with the differential $d_{CD'\bf g}$ of $CD'{\bf g}$.
(The differential of $D'\bf g$, which we denote $d_{D'\bf g}$, does \emph{not} participate
     in these definitions, but will play its role.)

For every   $x\in D'\bf g$ we denote ${\cal L}\langle x\rangle$ and $\iota\langle x\rangle$
the corresponding elements of $CD'\bf g$, and ${\cal L}_W\langle x\rangle$ and $\iota_W\langle x\rangle$
their action in $W$. With the notations of  Eqs. (\ref{EmbeddingL}) and (\ref{lF}),
Eq. (\ref{DMwithExpIota}) implies, for all $\omega\in W$:
\begin{align}  
 &(d_W + \iota_W\langle l \langle F\rangle\rangle) \exp\left(-\iota_W\langle i(F)\rangle\right) \omega \;=\;\nonumber{} \\ \;=\;
 &\exp\left(
                     -\iota_W\langle i(F)\rangle
                     \right)
                    \left(d_W + \iota_W\langle d_{D'\bf g}i(F)\rangle + {\cal L}_W\langle i(F)\rangle\right)\omega
                    \label{EIota} \end{align}
Let us consider Eq. (\ref{CartanWithEIota}) in the special case when $\omega$ satisfies:
\begin{align}  
 &d_W\;\omega = 0\label{OmegaIsClosed} \\  
 &\left((-)^{\bar{x}+1}\iota_W\langle d_{D'\bf g}x\rangle + {\cal L}_W\langle x\rangle\right)\omega = 0\label{SpecialRelation} \end{align}
for all $x\in D'\bf g$. Then:
\begin{equation}
   (d_W + \iota_W\langle l\langle F\rangle\rangle) \exp\left(-\iota_W\langle i(F)\rangle\right) \omega = 0
   \end{equation}
Eq. (\ref{SpecialRelation}) is a special requirement on $W$ and $\omega$.
It is by no means automatic. Intuitively, it may be understood as an interplay between $d_W$ and $d_{D'\bf g}$
(and $\omega$):
\begin{align}  
 &d_W\iota_W\langle x\rangle \omega = \iota_W\langle d_{D'\bf g}x\rangle\omega\label{InterplayD} \end{align}
We do not requite that $d_{D'\bf g}$ act in $W$.
Instead, we want Eqs. (\ref{OmegaIsClosed}) and (\ref{SpecialRelation})
(or, equivalently, Eqs. (\ref{OmegaIsClosed}) and (\ref{InterplayD})).

We will now consider two examples of $D'\bf g$-differential modules.

\subsection{Pseudo-differential forms (PDF)}\label{sec:SpecialPDFs}

Suppose that a supermanifold $M$ comes with an infinitesimal action of $D'{\bf g}$,
\textit{i.e.} a homomorphism:
\begin{equation}
   r\;:\;{D'\bf g} \rightarrow \mbox{Vect}(M)
   \label{FromDgToVect}\end{equation}
This is only a homomorphism of Lie superalgebras;
we forget, for now, about the differential $d_{D'\bf g}$.

We denote $d_M$ the deRham differential on $M$, and $F = F^a e_a$.
Then Eq. (\ref{DMwithExpIota}) implies 
\begin{equation}
   (d_M + \iota\circ F) \exp(-\iota\circ\alpha) \omega =
   \exp(-\iota\circ\alpha) (d_M + \iota\circ (d_{D'\bf g}\alpha) + {\cal L}\circ\alpha)\omega
   \label{CartanWithEIota}\end{equation}
Let us consider Eq. (\ref{CartanWithEIota}) in the special case when $\omega$ is closed:
\begin{equation}
   d\omega = 0
   \end{equation}
Consider a linear subspace ${\cal X}_{\omega}\subset \mbox{Vect}(M)$ consisting of all vectors $v$ such that exits
some other vector $d_{\omega}v\in {\cal X}_{\omega}$ satisfying: 
\begin{equation}
   - d\iota\langle v\rangle \omega = \iota\langle d_{\{\omega\}} v\rangle\omega
   \label{DefDOmega}\end{equation}
Suppose that $\omega$ is ``non-degenerate'' in the sense that the map from
$\mbox{Vect}M$ to PDFs on $M$ given by $v\mapsto \iota_v\omega$ is injective.
Then Eq. (\ref{DefDOmega}) defines an odd nilpotent operator:
\begin{equation}
   d_{\{\omega\}} \;:\; {\cal X}_{\omega} \rightarrow {\cal X}_{\omega}
   \end{equation}
Moreover, ${\cal X}_{\omega}$ is closed under the operation of commutator of vector fields.
Indeed:
\begin{align}  
 &(-)^{v+w+1}d\iota_{[v,w]}\omega = {\cal L}\langle [v,w]\rangle \omega = (-)^{v+w+1} d\iota\langle [v,w]\rangle\omega \;=\;\nonumber{} \\ \;=\;
 &(-)^w {\cal L}\langle v\rangle\iota\langle d_{\{\omega\}}w\rangle \omega
           + (-)^{vw+v+1} {\cal L}\langle w\rangle \iota\langle d_{\{\omega\}}v\rangle \omega \;=
           \nonumber{} \\ \;=\;
 &\iota\langle (-)^w[v,d_{\{\omega\}}w] + (-)^{vw + v + 1} [w,d_{\{\omega\}}v]\rangle\omega\nonumber{} \end{align}
Therefore $({\cal X}_{\omega},d_{\{\omega\}})$ is a differential Lie superalgebra.

Suppose that:
\begin{align}  
 &\mbox{im} \left(r\,:\,{D'\bf g}\rightarrow \mbox{Vect}(M)\right) \;\subset\; {\cal X}_{\omega}\nonumber{} \\  
 &d_{\{\omega\}}r(x) = r(d_{D'\bf g}x)\nonumber{} \end{align}
Then Eq. (\ref{CartanWithEIota}) implies that:
\begin{align}  
 &(d_M + \iota\circ F) \exp(\iota\circ\alpha) \omega = 0\nonumber{} \end{align}
In other words, $\omega^{\tt C}$ defined by the equation:
\begin{equation}
   \omega^{\tt C} =  \exp(\iota\circ\alpha) \omega
   \end{equation}
is a cocycle in the Cartan's model of equivariant cohomology.

Notice that apriori there is no action of $d_{D'\bf g}$ on $M$, and we have never used it.

\subsection{Special cocycles}\label{sec:SpecialCocycles}

Similarly, suppose that $D'\bf g$ is mapped into some Lie superalgebra $\bf a$:
\begin{equation}
   r\;:\;{D'\bf g} \rightarrow {\bf a}
   \end{equation}
and $W$ is a representation of $\bf a$. 
Consider the Chevalley-Eilenberg cochain complex $C({\bf a},W)$  of $\bf a$ with coefficients in $W$.
The cone $C\bf a$ acts on  $C({\bf a},W)$;
for each $x\in \bf a$ we denote ${\cal L}\langle x\rangle$ and $\iota\langle x\rangle$
the action of the corresponding elements of $C\bf a$.
Eq. (\ref{CartanWithEIota}) still holds, now $\omega$  is a cochain:
\begin{equation}
   \omega\in C({\bf a},W)
   \end{equation}
We are allowing arbitrary dependence of cochains on the ghosts of $\bf a$, not only polynomials.
We will say that a cocycle $\omega$ is \emph{special} if exists $d_{\{\omega\}}$ satisfying Eq. (\ref{DefDOmega})
for all $v\in\mbox{im}(r)$.

Moreover, we require:
\begin{equation}
   d_{\{\omega\}} \circ r = r \circ d_{D'\bf g}
   \label{RIntertwinesDifferentials}\end{equation}

\subsection{A procedure for constructing $r$}\label{sec:ProcedureForR}

We will now describe a procedure for constructing an embedding:
\begin{equation}
   r\;:\;{D\bf g} \rightarrow {\bf a}
   \end{equation}
This is not really an ``algorithm'' because, as we will see,  it may fail at any step.

Suppose that we can choose, for each $a\in \{1,\ldots,\mbox{dim} {\bf g}\}$, some
$\phi_a\in \bf a$ so that exist $\phi_{ab}\in{\bf a}$ such that:
\begin{align}  
 &[ \phi_a , \phi_b ] = d \phi_{ab}\nonumber{} \\  
 &[ \phi_a , d \phi_b ] = f_{ab}{}^c \phi_c\nonumber{} \end{align}
where $d=d_{\omega}$ and $f_{ab}{}^c$ the structure constants of $\bf g$. Then verify the existence
of $\phi_{abc}$ such that:
\begin{align}  
 &[\phi_a,\phi_{bc}] + [\phi_{ab},\phi_c] = d\phi_{abc}\nonumber{} \\  
 &[\phi_{ab},d\phi_c] = f_{ac}{}^d \phi_{db} + f_{bc}{}^d\phi_{ad}\nonumber{} \end{align}
The mutual consistency of these two equations follows from Eq. (\ref{DsAnticommute}).
Then continue this procedure order by order in the number of indices:
\begin{align}  
 &[\phi_{a_1},\phi_{a_2\ldots a_n}] + \ldots + [\phi_{a_1\ldots a_{n-1}},\phi_{a_n}] =
                                          d\phi_{a_1\ldots a_n}\nonumber{} \\  
 &[\phi_{a_1\ldots a_{n-1}},d\phi_b] = f_{a_1 b}{}^c \phi_{c a_2\ldots a_{n-1}} + \ldots\nonumber{} \end{align}
If we are able to satisfy these equalities, order by order in the number of indices, then we can put,
in notations of Eqs. (\ref{iF}) and (\ref{lF}):
\begin{align}  
 &r\langle l_a\rangle = d\phi_a\nonumber{} \\  
 &r\langle i_{a_1\ldots a_n}\rangle = \phi_{a_1\ldots a_n}\nonumber{} \end{align}

\section{Chevalley-Eilenberg complex of a differential module}\label{sec:ExpOfIota}

         In this Section, there is no $D{\bf g}$ nor $D'\bf g$. We forget about them for now.
As a preparation for BV formalism, we will now discuss another formula similar to Eq. (\ref{DWithExp}).

Consider a Lie superalgebra $\bf a$, its universal enveloping algebra $U\bf a$, and its cone $C{\bf a}$, generated by
${\cal L}\langle x\rangle$ and $\iota\langle x\rangle$,  $x\in \bf a$.
We consider the quadratic-linear dual coalgebra $U{\bf a}^{\ashriek}$.
The dual space $\left(U{\bf a}^{\ashriek}\right)^* = \mbox{Hom}\left(U{\bf a}^{\ashriek},{\bf K}\right)$
is the algebra of functions of the ``ghost variables'' $c^A$.
Following Section 3.4 of \cite{LodayVallette}, the Koszul twisting morphism is:
\begin{equation}
   \kappa = c^A e_A\;:\; U{\bf a}^{\ashriek} \rightarrow {\bf a} \subset U{\bf a}
   \label{KoszulTwistingMorphism}\end{equation}
We will study the properties of the following operator:
\begin{align}  
 &\exp\left(\iota\circ\kappa\right)\;\in\; \mbox{Hom}\left(U{\bf a}^{\ashriek}, UC{\bf a}\right)\nonumber{} \\  
 &\iota\circ\kappa = c^A\iota\langle e_A\rangle = \iota\langle c\rangle\nonumber{} \end{align}
Since $\bold a$ is quadratic-linear, $U{\bf a}^{\ashriek}$ comes with the differential
$d_{U{\bf a}^{\ashriek}}$. The dual differential on $\left(U{\bf a}^{\ashriek}\right)^*$
is the \emph{Chevalley-Eilenberg differential} $d_{\rm CE}^{(0)}$ (the BRST operator):
\begin{equation}
   d_{\rm CE}^{(0)} = \left(d_{U{\bf a}^{\ashriek}}\right)^* = f^A_{BC} c^Bc^C{\partial\over\partial c^A}
   \end{equation}
Here $d_{U{\bf a}^{\ashriek}}$ is the ``internal'' differential of $U{\bf a}^{\ashriek}$;
it comes from $U{\bf a}$ being inhomogenous
(\textit{i.e.} quadratic-linear and not purely quadratic algebra).

The Chevalley-Eilenberg complex $C^{\bullet}({\bf a}, W)$ with coefficients in $W$ can be defined
for any $\bf a$-module $W$. Consider the special case when $W$ is a $\bf a$-differential
module $W$, \textit{i.e.}
a representation of $(C{\bf a},d_{C{\bf a}})$. We will denote
${\cal L}_W\langle x\rangle$, $\iota_W\langle x\rangle$ and $d^W_{C\bf a}$ the elements of
$\mbox{End}(W)$ representing elements ${\cal L}\langle x\rangle$,  $\iota\langle x\rangle$
of $C\bf a$ and $d_{C\bf a}$.
(Then $W$ is also a representation
of $\bf a$, where $x\in\bf a$ is represented by ${\cal L}_W\langle x\rangle$.)

Consider the Lie algebra cochain complex (= Chevalley-Eilenberg complex)
of $\bf a$  with coefficients in
$W$. The differential is defined as follows:
\begin{align} d_{\rm CE}\;:\;
 &\mbox{Hom}\left(U{\bf a}^{\ashriek},W\right) \rightarrow \mbox{Hom}\left(U{\bf a}^{\ashriek},W\right)\nonumber{} \\ d_{\rm CE} \;=\;
 &\left(d_{U{\bf a}^{\ashriek}}\right)^* + {\cal L}_W\circ \kappa
                \;=\;d_{\rm CE}^{(0)} + {\cal L}_W\langle c\rangle\nonumber{} \\ d^{(0)}_{\rm CE} \;=\;
 &f^A_{BC} c^Bc^C{\partial\over\partial c^A}\nonumber{} \end{align}
All this can be defined for any  $\bf a$-module $W$. But when $W$ is also an $\bf a$-\emph{differential}
module (\textit{i.e.} a represenatation of $({C\bf a},d_{C\bf a})$), then $d_{\rm CE}$ and $d_{\rm CE}^{(0)}$ are related:
\begin{equation}
     (d^W_{C\bf a} + d_{\rm CE})\exp\,\iota_W\circ\kappa
     = \left(\exp\,\iota_W\circ\kappa \right)(d^W_{C\bf a} + d_{\rm CE}^{(0)})
   \label{ExpIota}\end{equation}
where $\iota_W\circ\kappa = \iota_W\langle c\rangle$. Notice that Eq. (\ref{ExpIota}) resembles Eq. (\ref{DWithExp}).
Indeed, both $\kappa$ of Eq. (\ref{KoszulTwistingMorphism}) and $\alpha$ of Eq. (\ref{NaturalTwistingMorphism})
are maps from coalgebra to algebra, satisfying the Maurer-Cartan (MC) equation.
But the way MC equation is satisfied is different,
because  $d_{\rm CE}^{(0)}$ acts in the coalgebra (in $Ua^{\ashriek}$)
while $d_{\Omega}$ acts in the algebra (in $\Omega(A^{\ashriek})$), see Section \ref{AsFaddeevPopov}.

Since $W$ and $U{\bf a}^{\ashriek}$ are both $C\bf a$-modules,
we can consider $\mbox{Hom}(U{\bf a}^{\ashriek}, W)$ a $C\bf a$-module, as a $\mbox{Hom}$ of
two $C\bf a$-modules:
\begin{align} \iota^{(0)}_{{\rm Hom}(U{\bf a}^{\ashriek}, W)}\langle x\rangle\;=\;
 &\iota_W\langle x\rangle + x^A{\partial\over\partial c^A}\nonumber{} \\ {\cal L}^{(0)}_{{\rm Hom}(U{\bf a}^{\ashriek}, W)}\langle x\rangle\;=\;
 &{\cal L}_W\langle x\rangle + x^Ac^Bf_{AB}^{C}{\partial\over\partial c^C}\nonumber{} \end{align}

\emph{Proposition \refstepcounter{Theorems}\label{BRSTofDifferentialModule}\noindent{\bf \arabic{Theorems}}}

\begin{align}  
 &\iota_{{\rm Hom}(U{\bf a}^{\ashriek}, W)}\langle x\rangle 
              \left(\exp\,\iota_W\circ\kappa\right) \;=\;
              \left(\exp\,\iota_W\circ\kappa\right)
              \iota^{(0)}_{{\rm Hom}(U{\bf a}^{\ashriek}, W)}\langle x\rangle \nonumber{} \\  
 &{\cal L}_{{\rm Hom}(U{\bf a}^{\ashriek}, W)}\langle x\rangle 
                \left(\exp\,\iota_W\circ\kappa\right) \;=\;
                \left(\exp\,\iota_W\circ\kappa\right)
                {\cal L}^{(0)}_{{\rm Hom}(U{\bf a}^{\ashriek}, W)}\langle x\rangle \nonumber{} \\ \mbox{where \hspace{2.00000ex}}
 &\iota_{{\rm Hom}(U{\bf a}^{\ashriek}, W)}\langle x\rangle =
                    x^A{\partial\over\partial c^A}\nonumber{} \\  
 &{\cal L}_{{\rm Hom}(U{\bf a}^{\ashriek}, W)}\langle x\rangle =
                {\cal L}^{(0)}_{{\rm Hom}(U{\bf a}^{\ashriek}, W)}\langle x\rangle\nonumber{} \end{align}
In other words, 

\noindent{}
both
$({\cal L}^{(0)}_{{\rm Hom}(U{\bf a}^{\ashriek}, W)},
            \iota^{(0)}_{{\rm Hom}(U{\bf a}^{\ashriek}, W)},
            d^W_{C\bf a} + d_{\rm CE}^{(0)})$

\noindent{}
and
$({\cal L}_{{\rm Hom}(U{\bf a}^{\ashriek}, W)},
            \iota_{{\rm Hom}(U{\bf a}^{\ashriek}, W)},
            d^W_{C\bf a} + d_{\rm CE})$
define on ${\rm Hom}(U{\bf a}^{\ashriek}, W)$ the structure of a differential $\bf a$-module,
and $\exp\,\iota_W\circ\kappa$ intertwines them.

\section{Integration measures from representations of $C\bf g$ and $D\bf g$}\label{PDFsFromRepresentations}

\subsection{PDFs from representations of $C\bf g$}\label{sec:PDFsFromCg}

If $\bf g$ acts on a manifold, then $C\bf g$ acts in PDFs. More generally, $C\bf g$
acts in cochains of Chevalley-Eilenberg complexes of $\bf g$.

\noindent{}
\emph{Question:} Given some representation $W$ of $(C{\bf g}, d_{C\bf g})$, can we map it to PDFs, or to cochains?

\subsubsection{Mapping to cochains}\label{sec:ToCochains}

\emph{Proposition \refstepcounter{Theorems}\label{DtoQ}\noindent{\bf \arabic{Theorems}}.}
Let $\cal W$ be a $\bf g$-module, and $\intoper : W\rightarrow \cal W$ an intertwiner of $\bf g$-modules:
\begin{align} \intoper\;:\;
 &W\rightarrow {\cal W}\nonumber{} \\ \mbox{satisfying: \hspace{2.00000ex}}
 &\intoper\circ d^W_{C\bf g} =0\label{IdCgZero} \end{align}
(One may think of $\intoper$ as an ``integration operation''.)
Consider the subspace
\begin{equation}
   \mbox{ker}\, d^{(0)}_{\rm CE} \subset \mbox{Hom}\left(U{\bf g}^{\ashriek},W\right)
   \end{equation}
Then operation $\intoper\circ e^{\iota_W\circ\kappa}$ intertwines this subspace
with the  Chevalley-Eilenberg complex $C^{\bullet}({\bf g}, {\cal W})$:
\begin{equation}
   \intoper\circ e^{\iota_W\circ\kappa}\;:\;
   (\mbox{ker} \,d^{(0)}_{\rm CE}\,,\, d^W_{C\bf g})
   \rightarrow
   (\mbox{Hom}\left(U{\bf g}^{\ashriek},{\cal W}\right)\,,\,d_{\rm CE})
   \end{equation}
\emph{Proof} follows from Eq. (\ref{ExpIota}).

Therefore every $w\in W$ defines an inhomogeneous Chevalley-Eilenberg cochain of $\bf g$ with coefficients in $\cal W$:
\begin{align}  
 &\psi\langle w\rangle  = \intoper e^{\iota_W\circ \kappa} w\nonumber{} \end{align}
The map    $\intoper\circ e^{\iota_W\circ\kappa}$ intertwines the action of $C\bf g$ on
$\mbox{ker} \,d^{(0)}_{\rm CE}$ with the standard action of $C\bf g$ in cochains with
coefficients in $\cal W$ --- the
$({\cal L}^{(0)}_{{\rm Hom}(U{\bf b}^{\ashriek}, {\cal W})},
            \iota^{(0)}_{{\rm Hom}(U{\bf b}^{\ashriek}, {\cal W})})$ of
\emph{Proposition} \ref{BRSTofDifferentialModule}.
(This action does not use $\iota^{\cal W}\langle x\rangle$,
      generally speaking there is no such thing as $\iota^{\cal W}\langle x\rangle$.
      Our $\cal W$, unlike $W$, is just a $\bf g$-module, not a differential $\bf g$-module.)

\subsubsection{Mapping to PDFs}\label{sec:ToPDFs}

Suppose that $\cal W$ happens to be a space of functions  on some manifold
$M$ with an action of  $G$ (the Lie group corresponding to $\bf g$).
In this case, every $w\in W$ and a point $m\in M$ defines a closed PDF on $G$, in the following way:
\begin{equation}
   \Omega\langle w\rangle(g,dg)=\left(\intoper e^{-\iota_W\langle dgg^{-1}\rangle} w\right)(g.m)
   \label{DefOmegaWI}\end{equation}
We will be mostly interested in the cases when this PDF descends to the $G$-orbit of $m$. 

For example, consider the case when $W$ is the space of PDFs on $M$ (the same $M$)
and $\intoper$ is the restriction of a PDF on the zero section $M\subset \Pi TM$.
(Remember that PDFs are functions on $\Pi TM$. In this example, the operation $\intoper$ associates
          to every form its 0-form component.) In this case, given a PDF on $M$,
\textit{e.g.} $f_{\mu}(x)dx^{\mu}$, our procedure, for each $x\in M$, associates to it a PDF on $G$,
which is just $f_{\mu}(g.x)d(g.x)^{\mu}$. 
If $G$ acts freely, $\Omega\langle w\rangle$ will descend to a form on the orbit of $x$. This is just the
restriction to the orbit of the original form we started with.

As another example, consider $\bf g$ the Lie algebra of vector fields on some manifold $N$,
and $W$ the space of PDFs on $N$. Let $M$ be the space of orientable $p$-dimensional
submanifolds of $N$, and $\intoper$ the operation of integration over such a submanifold.
Our construction maps closed forms on $N$ to closed forms on $M$.

\subsection{PDFs from representations of $D\bf g$}\label{PDFsFromDg}

An analogue of Eq. (\ref{ExpIota}) holds for $D\bf g$. It follows as a particular case from
the results of \cite{Alekseev:2010gr}:
\begin{equation}
   \left(d_{D\bf g} + d_{\rm CE}\right) \Phi = \Phi \left(d_{D\bf g} + d^{(0)}_{\rm CE}\right)
   \end{equation}
where $\Phi$ is:
\begin{align}  
 &\Phi = P\exp\int_0^1  {\cal A}_{\tau}d\tau\nonumber{} \\  
 &A_{\tau}d\tau = \left.{d\over du}\right|_{u=0}i(u \, d\tau \, c   + (\tau - \tau^2)c^2)\nonumber{} \end{align}
Therefore, when $W$ is a representation of $D\bf g$, we have an analogue of Eq. (\ref{DefOmegaWI}),
where $\intoper$ should now satisfy $\intoper\circ d_{D\bf g}^W = 0$:
\begin{align}  
 &\Omega\langle w\rangle = \left(\intoper\; \Phi|_{c \mapsto dg g^{-1}}\,w\right)(g.m)\label{OmegaFromDg} \\  
 &d\Omega \langle w\rangle = \Omega\langle d^W_{D\bf g}w\rangle\nonumber{} \end{align}

\section{BV}\label{BV}

We will now apply the technique developed in the previous sections to the BV formalism.

Let $\bf a$ be the Lie algebra of functions on the BV phase space with flipped statistics.
Its elements are $s^{-1}f$ where $f$ is a function on the BV phase space and $s$ the suspension:
\begin{equation}
   s^{-1}f\in \bf a
   \end{equation}
The Lie bracket is given by the odd Poisson bracket.

\subsection{Half-densities as a representation of $C{\bf a}$}\label{sec:BVRepOfCone}

The space of half-densities on the BV phase space is a representation of $C{\bf a}$ :

\begin{align} d \rho_{1/2} =
 &\;  - \Delta_{\rm can} \rho_{1/2}\nonumber{} \\ \iota\langle s^{-1}f\rangle \rho_{1/2} =
 &\;  f\rho_{1/2}\nonumber{} \\ {\cal L}\langle s^{-1}f\rangle \rho_{1/2} =
 &\; (-)^{|f|}\Delta_{\rm can} (f\rho_{1/2}) - f\Delta_{\rm can}  \rho_{1/2}\nonumber{} \end{align}

We are now in the context of  Section \ref{sec:PDFsFromCg}.
Now $\bf g$ is $\bf a$, $W$ is the space of half-densities,
$d^W_{C\bf g}$ is $-\Delta_{\rm can}$
and $\cal W$ is $\mbox{Fun}({\rm LAG})$ --- the space of functions on Lagrangian submanifolds.

\subsection{Correlation functions as a Lie superalgebra cocycle}\label{sec:CorrelationFunctionsAreCocycle}

Correlation function defines a linear map:
\begin{align}  
 &U{\bf a}^{\ashriek} \longrightarrow \mbox{Fun}({\rm LAG})\nonumber{} \\  
 &f_1\bullet\cdots\bullet f_n \mapsto \left[L \mapsto \int_L f_1\cdots f_n \rho_{1/2}\right]\label{BVCocycle} \end{align}
where $\bullet$ means symmetrized tensor product
(examples of sign rules are in Section \ref{sec:DualAlgebraOfFunctions}).

\emph{Proposition \refstepcounter{Theorems}\label{CorrelatorAsCocycle}\noindent{\bf \arabic{Theorems}}}
Eq. (\ref{BVCocycle}) defines an injective map from the space of half-densities to the space of
cochains of $\bf a$ with values in functionals on Lagrangian submanifolds;
to every half-density $\rho_{1/2}$ corresponds a cochain given by Eq. (\ref{BVCocycle}).
This map is an intertwiner of the actions of the differential Lie superalgebra $C{\bf a}$.
In particular, if $\rho_{1/2}$ satisfies the Quantum Master Equation:
\begin{equation}
   \Delta_{\rm can} \rho_{1/2} = 0
   \end{equation}
then Eq. (\ref{BVCocycle}) defines a cocycle of $\bf a$ with coefficients
in the space of functionals on Lagrangian submanifolds.

\emph{Proof} follows from Proposition \ref{DtoQ}.

\noindent{}
The image of this map consists of the cochains satisfying the following \emph{locality property}.
Given $f_1,\ldots,f_n$, if for some $i$ and $j$  $\mbox{supp}(f_i)\cap \mbox{supp}(f_j) = \mbox{\tt empty set}$
then $c(f_1\bullet\ldots\bullet f_n) = 0$.
It is important for us, that this subset is preserved by the canonical transformations,
\textit{i.e.} by the action of $\bf a$ on its cocycles.

Cocycles with
coefficients in $\mbox{Fun}({\rm LAG})$, defined by Eq. (\ref{BVCocycle}), 
can be interpreted as closed differential forms on $\rm LAG$, by the construction
of Section \ref{sec:ToPDFs}. We take:
\begin{align}  
 &M=\rm LAG\label{MisLAG} \\  
 &\intoper = \left[L\mapsto \int_{L\in\rm LAG} \_ \right]\label{IntoperLAG} \end{align}

\section{Equivariant BV formalism}\label{EquivariantBV}

\subsection{Equivariantly closed cocycle in the Cartan model}\label{sec:EquivariantCocycle}

For all cocycles coming from half-densities, Eq. (\ref{DefDOmega}) is satisfied with:
\begin{align}  
 &d_{\{\omega\}} f = \Delta_{\rho_{1/2}} f\nonumber{} \\ \mbox{where \hspace{1.00000ex}}
 &\Delta_{\rho_{1/2}}f = \rho_{1/2}^{-1}\Delta_{\rm can} \left(f\rho_{1/2}\right)\nonumber{} \end{align}
As in  Section \ref{sec:SpecialCocycles},
suppose that $r$ is an embedding of $D'\bf g$ in $\bf a$.
Eq. (\ref{RIntertwinesDifferentials}) becomes (\textit{cp} Eqs. (\ref{iF}) and (\ref{lF})):
\begin{align}  
 &\Delta_{\rho_{1/2}} r\langle i(F)\rangle + {1\over 2}[r\langle i(F)\rangle,r\langle i(F)\rangle] =
               r\langle l\langle F\rangle\rangle\label{DeltaR} \\  
 &[r\langle l\langle F_1\rangle\rangle , r\langle i(F_2)\rangle]
        = \left.{d\over dt}\right|_{t=0} r\langle i(e^{t[F_1,\_]}F_2)\rangle\label{Equivariance} \\  
 &\mbox{where $i(F)$ and $l\langle F\rangle$ were defined in Eqs. (\ref{iF}) and (\ref{lF})}\nonumber{} \end{align}
Then, equivariantly closed cocycle in the Cartan model is given by:
\begin{equation}
   f_1\bullet\cdots\bullet f_n \mapsto
   \left[L \mapsto \int_L f_1\cdots f_n e^{r\langle i(F)\rangle}\rho_{1/2}\right]
   \end{equation}
Our notations here differ from our previous papers;
$r\langle i(F)\rangle$ was called $\Phi(F)$ in Section 4 of \cite{Mikhailov:2016myt}
and $a(F)$ in Section 6 of \cite{Mikhailov:2016rkp}. Here is the summary of notations:

\begin{tabular}{|l|l|l|l|l|l|l|}\hline here&$F$&$i(F)$&$l\langle F\rangle$&$r\langle i(F)\rangle$&$r\langle i_aF^a\rangle$&$r\langle l\langle F\rangle\rangle$\\\hline  \cite{Alekseev:2010gr}&$t$&$i(t)$&$l(t)$&&&\\\hline  Section 4 of \cite{Mikhailov:2016myt}&$h$&&&$\Phi(h)$&&\\\hline  Section 6 of \cite{Mikhailov:2016rkp}&$\xi$&&&$a(\xi)$&$\Phi\langle \xi\rangle$&$\underline{\xi}$\\\hline \end{tabular}

\subsection{Deformations}\label{sec:Deformations}

If $r\langle l\langle F\rangle\rangle$ and $r\langle i(F)\rangle$ solve
Eqs. (\ref{DeltaR}) and (\ref{Equivariance}) with some $\rho_{1/2}$, then
$r\langle l\langle F\rangle\rangle$ and $r\langle i(F)\rangle - f$ solve
with $e^f\rho_{1/2}$. This allows us to assume, without loss of generality, that $r\langle i(0)\rangle =0$,
\textit{i.e.} to consider representations of $D\bf g$ rather than $D'\bf g$. 
Or, we can fix $\rho_{1/2} = \rho_{1/2}^{(0)}$ for some fixed $\rho_{1/2}^{(0)}$. Let us fix  the half-density,
delegating the deformations $\rho_{1/2}$ into $r\langle i(0)\rangle$.

Consider the deformations of the embedding $r\;:\; {D'\bf g} \rightarrow {\bf a}$ keeping
$r\langle l\langle F\rangle\rangle$ fixed.
\footnote{
         Notice that $r\langle l\langle F\rangle\rangle$ describes the action of symmetries on the BV
         phase space. We do not want to deform the action of symmetries.
         }
Eq. (\ref{DeltaR}) implies that a small variation $\delta r\langle i(F)\rangle$ satisfies:
\begin{align}  
 &\left(
              \Delta_{\rho_{1/2}} + 
              [r\langle i(F)\rangle,\_]
              \right)\;\delta r\langle i(F)\rangle \;=\; 0\nonumber{} \\  
 &[r\langle l\langle F_1\rangle\rangle , \delta r\langle i(F_2)\rangle]
        = \left.{d\over dt}\right|_{t=0} \delta r\langle i(e^{t[F_1,\_]}F_2)\rangle\nonumber{} \end{align}
Those $\delta r$ which are in the image of $\Delta_{\rho_{1/2}} + [r\langle i(F)\rangle,\_]$
correspond to trivial deformations. This means that 
the cohomologies of the operator $d_{D{\bf g}} + [i(F),\_]$, considered in \cite{Alekseev:2010gr},
in our context compute \emph{infinitesimal deformations} of the equivariant half-density.

\section{Integrating unintegrated vertices}\label{IntegratedAndUnintegrated}

\subsection{Integration prescription using $C\bf g$}\label{sec:IntegrateUsingCone}

Let us fix $i(0)=0$, and consider $r\,:\,D{\bf g}\rightarrow {\bf a}$.
In our discussion in Section \ref{sec:Deformations},
we assumed that deformations preserve the symmetries of $\omega$, \textit{i.e.} that $f$ is
$\bf g$-invariant. In string theory, it is useful to consider more general deformations
breaking $\bf g$ down to a smaller subalgebra ${\bf g}_0$. They are called ``unintegrated vertex operators''.
As their name suggests, $\bf g$-invariant deformations can be obtained by integration
over the orbits of $\bf g$. The procedure of integration
\href{https://andreimikhailov.com/math/bv/unintegrated-vertex/index.html}{\textbf{\textcolor{blue}{was described}}}
in \cite{Mikhailov:2016rkp}.
It is a particular case of Section \ref{sec:PDFsFromCg},
where $M$ is now $\rm LAG$ --- the space of Lagrangian submanifolds, and
$\intoper$ the operation of integration of half-density over a Lagrangian submanifold.

For this construction, we do not need the full $r\,:\, D{\bf g} \rightarrow {\bf a}$, but only
its restriction on ${\bf g}\subset {\bf a}$:
\begin{align}  
 &l\,:\,{\bf g}\rightarrow D{\bf g}\nonumber{} \\  
 &r\circ l\,:\,{\bf g}\rightarrow {\bf a}\nonumber{} \end{align}
For $v\in {\bf a}$, consider the deformations of $\omega$ of the following form:
\begin{equation}
   \delta\omega = \iota\langle v\rangle \omega
   \label{IotaDeformation}\end{equation}
In terms of half-densities:
\begin{equation}
   \delta \rho_{1/2} = \underline{v}\rho_{1/2}
   \end{equation}
where $\underline{v}$ denotes (as in \cite{Mikhailov:2016rkp}) the BV Hamiltonian generating $v$.

The cone $C\bf g$ acts on such deformations $\delta\rho_{1/2}$;
the action of $\hat{\iota}\langle x\rangle$, $\hat{\cal L}\langle x\rangle$
and $\widehat{d_{C\bf g}}$ is:
\begin{align}  
 &\hat{\iota}\langle x\rangle \delta\omega = -((\iota\circ r\circ l)x)\delta\omega\nonumber{} \\  
 &\hat{\cal L}\langle x\rangle \delta\omega = -(({\cal L}\circ r\circ l)x)\delta\omega\nonumber{} \\  
 &\widehat{d_{C\bf g}}\delta \omega = d\delta \omega\label{dCgActsAsDelta} \end{align}
Therefore the construction of Section \ref{sec:PDFsFromCg}, with $M=\rm LAG$
and $\intoper\langle \delta\omega\rangle = \left[L\mapsto \int_{L\in\rm LAG}f\rho_{1/2}\right]$,
gives a closed form on $G$ for every deformation of the form Eq. (\ref{IotaDeformation}):
\begin{align} \Omega\langle v\rangle
 &\;\in\;\mbox{Fun}(\Pi TG)\nonumber{} \\ \Omega\langle v\rangle
 &\;=\; 
           \Big(\intoper\;\exp(\iota\langle r\langle l\langle dg g^{-1}\rangle\rangle\rangle)\;
           \iota\langle v\rangle \omega\Big)(gL)\;=\;\nonumber{} \\  
 &\;=\; \int_{gL}\exp\left(\underline{r\langle l\langle dg g^{-1}\rangle\rangle}\right)
              \underline{v} \rho_{1/2}\nonumber{} \end{align}
This is just a particular case of the general construction of
Section \ref{sec:CorrelationFunctionsAreCocycle},
Eqs. (\ref{MisLAG}), (\ref{IntoperLAG}). We restrict the general construction of $\Omega$
from all $\rm LAG$ to an orbit of $G$. In other words, we consider not all odd canonical transformations,
but only a subgroup $G$.
But now we can use $G$-invariance of $\rho_{1/2}$ to pull back to a fixed $L$:
\begin{equation}
   \Omega\langle f\rangle
   = \int_{L}\exp\left(\underline{r\langle l\langle dg g^{-1}\rangle\rangle}\circ g\right)
   (\underline{v}\circ g) \rho_{1/2}
   \label{NonLocalOmega}\end{equation}
However Eq. (\ref{NonLocalOmega}) is somewhat unsatisfactory. 
Although it \emph{is}, actually, the integrated vertex corresponding to $v$,
this form of presenting it makes it apparently nonlocal on the string worldsheet.
We would want, instead, to replace, roughly speaking, $\iota{\cal L}$ with  ${\cal L}\iota$:
\begin{equation}
   \exp\left(\iota\langle r\langle l\langle dg g^{-1}\rangle\rangle\right)
   \stackrel{?}{\mapsto}
   \exp\left({\cal L}\langle r\langle i(dg g^{-1})\rangle\rangle\right)
   \end{equation}
We will now explain the construction.

\subsection{Integration using $D\bf g$}\label{sec:IntegrateUsingDg}
\subsubsection{Deformations as a representation of $D\bf g$}\label{sec:DeformationsAsRepOfDg}

In deriving Eq. (\ref{NonLocalOmega}) we have not actually used the representation of $D\bf g$,
but only the representation of ${\bf g}\subset D{\bf g}$;
we have only used $l\langle dg g^{-1}\rangle$ and never $i(dgg^{-1})$.
Notice, however, that the whole $D\bf g$ acts on deformations.
(This is, ultimately, due to our requirement of $\rho_{1/2}$ being ``equivariantizeable'',
      Section \ref{EquivariantBV}.)
Moreover, if we do not care about $d_{D\bf g}$, then there are two ways of defining the action of
just $D\bf g$. (It is easy to construct representations of free algebras.)
The first way is to use the embedding ${D\bf g}\stackrel{\cal L}{\rightarrow} CD{\bf g}$.
But this one does not define the action of $d_{D\bf g}$.

There is, however, the second way, which defines the action of $D{\bf g}$ \emph{with} its differential
$d_{D\bf g}$. For $\delta\omega =\iota\langle v\rangle \omega$ or equivalently $\delta \rho_{1/2} = \underline{v}\rho_{1/2}$, we define:
\begin{align}  
 &\hat{l}\langle F\rangle \;v \;=\; [ r\langle l\langle F\rangle\rangle\,,\,v]\nonumber{} \\  
 &\hat{i}(F) \;v \;=\; [ r\langle i(F)\rangle\,,\,v]\nonumber{} \\  
 &\widehat{d_{D\bf g}}\delta\omega = d\delta\omega\label{dDgActsAsDelta} \end{align}
To summarize, the space of deformations of $\omega$ can be considered
as a representation of $(CD{\bf g},d_{CD{\bf g}})$, or as a representation of $(D{\bf g},d_{D\bf g})$. (But not of $(CD{\bf g}, d_{CD{\bf g}}, d_{D\bf g})$ whatever that would be.)
In both cases, the differential acts as $\Delta$. That is to say,
the $d_{CD{\bf g}}$ of $(CD{\bf g},d_{CD{\bf g}})$ acts as $\Delta$,
and the $d_{D{\bf g}}$ of $(D{\bf g},d_{D{\bf g}})$ also acts as $\Delta$
--- see Eqs. (\ref{dCgActsAsDelta}) and (\ref{dDgActsAsDelta}), respectively.

\subsubsection{Averaging procedure using $D\bf g$}\label{sec:AveragingUsingDg}

Now we can apply the construction of Section \ref{PDFsFromDg}.
Eq. (\ref{OmegaFromDg}) gives:
\begin{align}  
 &\Omega\langle v\rangle = \int_{gL}
                      \left[\left(P\exp\int_0^1 {\cal A}_{\tau}d\tau\right) v\right]\rho_{1/2}\label{LocalOmega} \\ \mbox{where\hspace{2.00000ex}}
 &A_{\tau}d\tau = \left.{d\over du}\right|_{u=0}
                \hat{i}(u \, d\tau \, dgg^{-1}   + (\tau - \tau^2)(dgg^{-1})^2)\nonumber{} \end{align}
For completeness, we compare the notations in the table:

\begin{tabular}{|l|l|l|l|l|}\hline now&$dg g^{-1}$&$0$&$\iota\langle r\langle l\langle dgg^{-1}\rangle\rangle\rangle$&${\cal L}\langle r\langle i_a(dgg^{-1})^a\rangle\rangle$\\\hline  \cite{Alekseev:2010gr}&$\theta$&$t$&--&${\cal I}(\theta)$\\\hline  \href{https://andreimikhailov.com/math/bv/unintegrated-vertex/Descent_Procedure.html}{\textbf{\textcolor{blue}{Section 12}}} of \cite{Mikhailov:2016rkp}&$c$&$0$&$\Delta \Psi$&$\{\Psi,\_\}$\\\hline \end{tabular}

\noindent{}
There is no $CD\bf g$ in \cite{Alekseev:2010gr}, only $D\bf g$.

\subsection{Relation between two integration procedures}\label{sec:RelationBetweenIntegrations}

In the special case when $D\bf g$ reduces to $C\bf g$ (\textit{i.e.} $i_{a_1\ldots a_n} = 0$ for $n>1$), it
\href{https://andreimikhailov.com/math/bv/unintegrated-vertex/Descent_Procedure.html}{\textbf{\textcolor{blue}{was found}}}
in \cite{Mikhailov:2016rkp}, to the second order in the expansion
in powers of $dgg^{-1}$, that the two PDFs are different by an exact PDF on $G$.
It must be true in general. 

\subsection{Is integration form base with respect to $G_0$?}\label{sec:BaseG0}

Let $G_0\subset G$ be the stabilizer of $v$. Eq. (\ref{NonLocalOmega}) is not base with respect
to ${\bf g}_0 =\mbox{Lie} G$. But it can be made base, provided that one can extend $v$ to the solution
of the equation:
\begin{align}  
 &\Delta v(F_0) + [i(F_0),v(F_0)] = 0\nonumber{} \\  
 &\left.{d\over dt}\right|_{t=0} v(e^{t[\xi,\_]} F_0) = [\xi,v(F_0)]\nonumber{} \\  
 &v(0)=v\nonumber{} \end{align}
Then, the equivariant version of Eq. (\ref{NonLocalOmega}) is:
\begin{align} \Omega \;=\;
 &\int_{L}\exp\left(\underline{r\langle l\langle dg g^{-1}\rangle + i(F_0)\rangle}\circ g\right)
                (\underline{v(F_0)}\circ g) \rho_{1/2}\;=\nonumber{} \\ \;=\;
 &\int_L \left.{d\over d\varepsilon}\right|_{\varepsilon=0}
                  \exp\left[
                            \underline{r\langle l\langle dg g^{-1}\rangle \rangle}\circ g
                            + \left(
                                    \underline{r\langle i(F_0)\rangle} + \varepsilon \underline{v(F_0)}
                                    \right)\circ g
                            \right]\rho_{1/2}\nonumber{} \end{align}
and one can obtain a base form by choosing a connection.

We do not know the equivariant version of Eq. (\ref{LocalOmega}).
In the pure spinor formalism, it is very likely that the form given by Eq. (\ref{LocalOmega}) is already base,
because unintegrated vertex operator does not contain derivatives \cite{Flores:2019dwr}.

Notice that the PDF defined in Eq. (\ref{LocalOmega}) does not, generally, speaking, descend
to the orbit of $L$. In computing the average, the integration variable is $g$,  not  $gL$.
However, the integral does not depend on the choice of $L$ in the orbit.

\appendix{}

\section{Nilpotence of $d_{D\bf g}$}\label{AppendixNilpotence}

\subsection{Commutator of $D\bf g$}\label{sec:AppendixCommutatorDg}

The commutator of $D\bf g$ was defined in Section \ref{paragraph:CommDg}.
In particular, when considering a commutator of an element of  $\bf g$ and an element of $\mbox{FreeLie}\left(s^{-1} \overline{A^{\ashriek}}\right)$,
the following description is useful.
Consider the $UC\bf g$ -- the universal enveloping algebra of $C\bf g$, and its
dual coalgebra $UC{\bf g}^{\ashriek}$:
\begin{equation}
   UC{\bf g}^{\ashriek} = T^c\left(s({\bf g} \oplus s{\bf g})\right)
   \end{equation}
Consider the projector $p$:
\begin{align} p\;:\;
 &UC{\bf g}^{\ashriek}\rightarrow s{\bf g}\oplus A^{\ashriek}\label{ProjectorP} \end{align}
which is identity on $s{\bf g}\oplus A^{\ashriek}\;\subset\;T^c\left(s({\bf g} \oplus s{\bf g})\right)$
and zero on all tensors of rank $\geq 2$ containing at least one $sx \in s{\bf g}$.
This induces a map from $\Omega^1 \left(UC{\bf g}^{\ashriek}\right)$ to $D\bf g$ which we also
denote $p$:
\begin{equation}
   \Omega^1 \left(UC{\bf g}^{\ashriek}\right) =
   s^{-1}UC{\bf g}^{\ashriek}
   \stackrel{p}{\longrightarrow}
   {\bf g}\oplus s^{-1}A^{\ashriek} \subset D{\bf g}
   \label{LiftedProjectorP}\end{equation}
For any Lie superalgebra $\bf a$ let $\gamma$ denote the commutator map:
\begin{align}  
 &\gamma\;:\; {\bf a}\otimes {\bf a} \rightarrow {\bf a}\nonumber{} \\  
 &\gamma(v\otimes w) = [v,w]\nonumber{} \end{align}
In case of ${\bf a} = D{\bf g}$, we can consider $D{\bf g} \otimes D{\bf g}$ as
a subspace in $\Omega^2(UC{\bf g}^{\ashriek})$ using the projector $p$ of Eq. (\ref{LiftedProjectorP}):
\begin{equation}
   p\otimes p \;:\; \Omega^2(UC{\bf g}^{\ashriek}) \rightarrow D{\bf g} \otimes D{\bf g}
   \end{equation}
Then, the commutator on $D{\bf g}$ satisfies:
\begin{align}  
 &
        \gamma\;(p\otimes p)\; d_{\Omega(UC{\bf g}^{\ashriek})} \;s^{-1}b \;=\; - s^{-1} d_{UC{\bf g}^{\ashriek}}b
        \nonumber{} \\ \mbox{where\hspace{1.00000ex}}
 &b\;=\;s\xi\otimes a + (-)^{(\xi+1)a} a\otimes s\xi\;\in\;UC{\bf g}^{\ashriek}\nonumber{} \\  
 &\xi\in {\bf g}\;,\quad a\in A^{\ashriek}\nonumber{} \end{align}

\subsection{Nilpotence of $d_{D\bf g}$}\label{sec:AppendixNilpotence}

We will now prove that $d'$ anticommutes with $d_{\Omega}$:
\begin{equation}
   d'd_{\Omega} + d_{\Omega} d' = 0
   \label{AppendixDsAnticommute}\end{equation}
When $b\in \left(A^{\ashriek}\right)^{(\geq 2)}$,
by definition $d' s^{-1}b =0$. We must therefore check that $d' d_{\Omega} s^{-1}b=0$:
\begin{equation}
   d' d_{\Omega} s^{-1}b = \gamma (p\otimes p) d_{\Omega(UC{\bf g}^{\ashriek})} s^{-1} d_{C\bf g} b =
   - s^{-1} d_{UC{\bf g}^{\ashriek}}d_{C\bf g}b =
   s^{-1} d_{C\bf g} d_{UC{\bf g}^{\ashriek}} b = 0
   \end{equation}

\section{Bibliography}\label{Bibliography}

\section*{Acknowledgments}

This work was supported in part by FAPESP grant 2019/21281-4.


\def\cprime{$'$} \def\cprime{$'$}
\providecommand{\href}[2]{#2}\begingroup\raggedright\endgroup

\end{document}